\documentclass[a4paper,11pt]{article}
\usepackage{jinstpub} 
\usepackage{lineno}
\usepackage[separate-uncertainty=true]{siunitx}
\DeclareSIUnit\bar{bar}
\usepackage{hyperref}
\usepackage{subcaption}
\usepackage{multirow}

\title{\boldmath A novel cryogenic VUV spectrofluorometer for the characterization of wavelength shifters}
\author[1,a]{A. Leonhardt,\note{Corresponding author}}
\author[a]{M. Goldbrunner,}
\author[b]{B. Hackett}
\author[a]{and S. Schönert}

\affiliation[a]{Technical University of Munich, TUM School of Natural Sciences, Department of Physics,\\
James-Franck-Str. 1, 85748 Garching, Germany}
\affiliation[b]{Max Planck Institute for Physics,
\\Boltzmannstr. 8 85748 Garching, Germany}

\emailAdd{andreas.leonhardt@tum.de}

\abstract{
We present a novel cryogenic VUV spectrofluorometer designed to characterize wavelength shifters (WLS) crucial for experiments based on liquid argon (LAr) scintillation light detection. Wavelength shifters like 1,1,4,4-tetraphenyl-1,3-butadiene (TPB) or polyethylene naphthalate (PEN) are used in these experiments to shift the VUV scintillation light to the visible region. Precise knowledge of the optical properties of the WLS at liquid argon's temperature (\SI{87}{\kelvin}) and LAr scintillation wavelength (\SI{128}{\nano\meter}) is necessary to model and understand the detector response. The cryogenic VUV spectrofluorometer was commissioned to measure the emission spectra and relative wavelength shifting efficiency (WLSE) of samples between \SIrange{300}{87}{\kelvin} for VUV (\SIrange{120}{190}{\nano\meter}) and UV (\SI{310}{\nano\meter}) excitation. New mitigation techniques for surface effects on cold WLS were established. As part of this work, the TPB-based wavelength shifting reflector (WLSR) featured in the neutrinoless double-beta decay experiment LEGEND-200 was characterized. The WLSE was observed to increase by \SI{54 \pm 5}{\percent} from room temperature (RT) to \SI{87}{\kelvin}. PEN installed in LEGEND-200 was also characterized, and a first measurement of the relative WLSE and emission spectrum at RT and \SI{87}{\kelvin} is presented. The WLSE of amorphous PEN was found to be enhanced by at least \SI{37 \pm 4}{\percent} for excitation with \SI{128}{\nano\meter} and by \SI{52 \pm 3}{\percent} for UV excitation at \SI{87}{\kelvin} compared to RT.}

\arxivnumber{2311.15901} 

\proceeding{LIDINE 2023: LIght Detection In Noble Elements\\
  September 20-22\\
  Madrid, Spain}

\begin{document}
\maketitle
\flushbottom

\section{Introduction}
\label{sec:intro}
Liquid argon (LAr) is a widely used scintillating medium in dark matter WIMP detectors, such as DEAP-3600\cite{DEAP} and DarkSide-20k\cite{DarkSide}, and in neutrinoless double-beta decay experiments like LEGEND-200\cite{LEGEND} for background suppression. Ionizing particles in LAr induce vacuum-ultraviolet (VUV) scintillation light at \SI{128}{\nano\meter}. Most light detectors have low sensitivity in the VUV range, necessitating the use of wavelength shifters (WLS). The widely used 1,1,4,4-tetraphenyl-1,3-butadiene (TPB) is applied as a coating to optically inactive surfaces or photo-detectors like photo-multiplier tubes. An alternative is the scintillating and wavelength-shifting polymer polyethylene naphthalate (PEN), increasingly popular for its commercial availability as pellets, which can be injection molded into structural parts or extruded into thin films\cite{PENproduction}.\par
The LEGEND-200 experiment investigates neutrinoless double-beta decay of \textsuperscript{76}Ge using \SI{200}{\kilo\gram} of enriched high-purity germanium (HPGe) crystals, which serve as both the decay source and detectors\cite{LEGEND}. A  LAr volume, acting as a coolant and active shield, surrounds the HPGe detectors. The light collection system consists of light-guiding fibers surrounding the detectors, coated in TPB and coupled to silicon photomultipliers (SiPMs). The LAr detector's active volume is confined by a \SI{1.4}{\meter} diameter, \SI{3}{\meter} height wavelength shifting reflector (WLSR) consisting of a Tetratex\textregistered$~$lining coated in-situ with TPB\cite{SchwarzANNIMA}. The WLSR shifts LAr scintillation light and reflects light towards the central readout, enhancing the detection efficiency. The detector holder plates are made out of PEN, increasing the background suppression efficiency near the detectors. They scintillate when traversed by ionizing radiation and shift LAr scintillation light to the visible blue.\par
To understand and model the response of the LAr detector of LEGEND-200 and other experiments, the optical properties of the WLS must be known precisely. This includes, among others, the emission spectrum and the wavelength shifting efficiency. For LAr-based experiments, this means that the materials need to be characterized at \SI{87}{\kelvin} and with VUV light excitation, i.e., \SI{128}{\nano\meter}. The wavelength-shifting properties of organic WLS are expected to change significantly with decreasing temperature due to the decline of vibrational and rotational relaxations in competition with radiative de-excitations\cite{Marcin}.\par
\section{Cryogenic VUV spectrofluorometer}
\label{sec:setup}
\begin{wrapfigure}{r}{0.48\textwidth}
  \centering
    \includegraphics[width=0.45\textwidth]{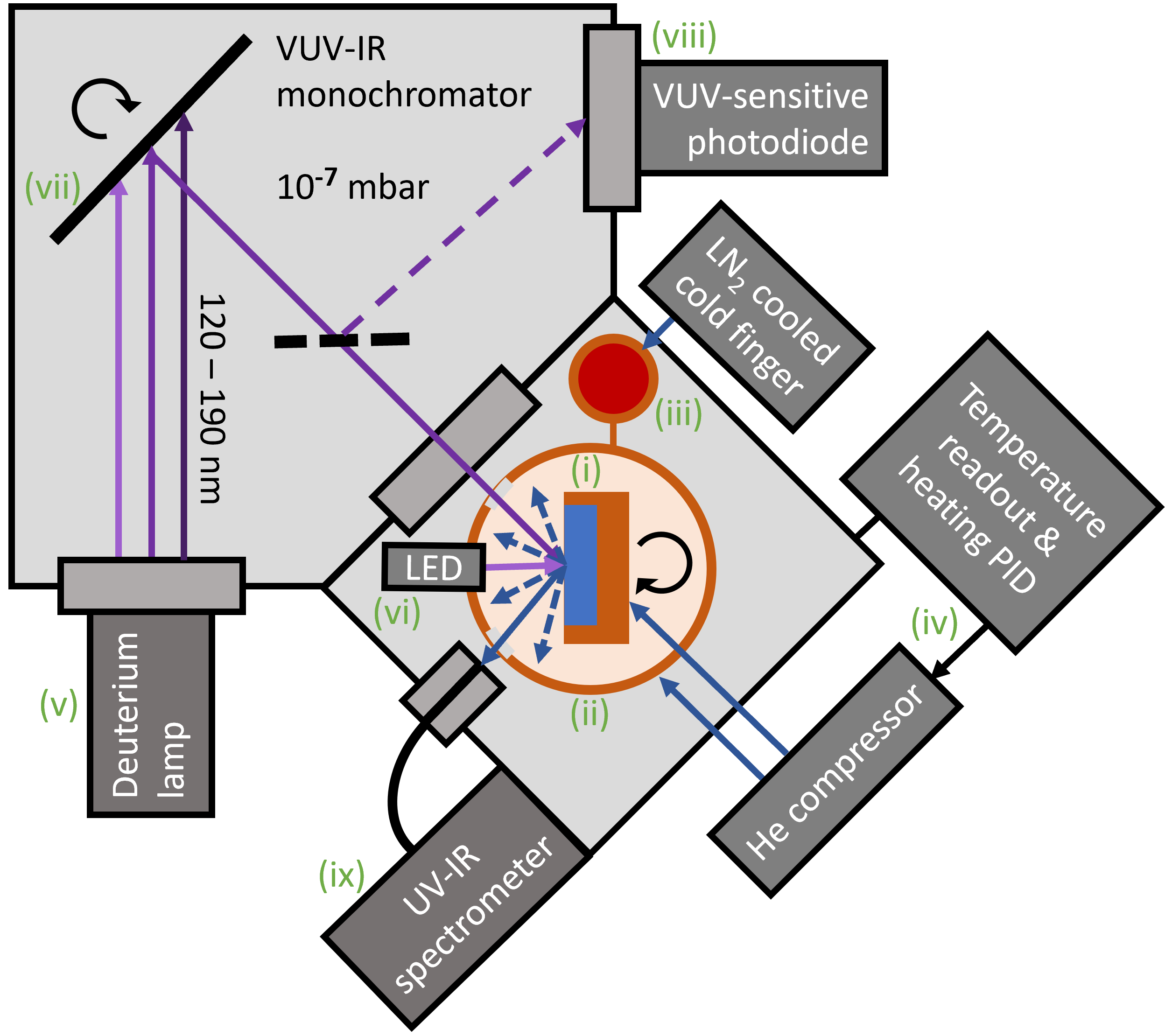}
    \caption{Schematic drawing of the VUV spectrofluorometer. The Roman numerals indicate components referenced in the text.}
    \label{fig:schematic}
\end{wrapfigure}

A dedicated experimental setup was designed and commissioned to characterize WLS at temperatures ranging from \SIrange{300}{87}{\kelvin}, with excitation wavelengths down to \SI{128}{\nano\meter}. The VUV spectrofluorometer is schematically depicted in \autoref{fig:schematic}.\par
In the sample chamber, a copper holder accommodates a $30 \times 30 \, \text{mm}^2$ WLS sample and is equipped with a \SI{30}{\watt} polyimide heater on its back (indicated in \autoref{fig:schematic} as (i)). The holder, positioned on the second stage of a two-stage Gifford McMahon cold head, is encased by a copper cryo-shield (ii), leaving openings for excitation light, light detection, and cabling. A copper beaker filled with silica gel is connected to a copper cold finger, which is partially submerged in a liquid nitrogen dewar outside the chamber (iii). The temperature at the sample holder recorded by two Pt-100 sensors is the input for a PID controlling the heater's power (iv)\cite{MaxiBA}. \par
A Hamamatsu L15094 deuterium lamp ($\lambda_{\text{Lamp}}=\SI{120}{\nano\meter}-\SI{190}{\nano\meter}$) (v) and a Marktech Optoelectronics MTE310H33-UV LED ($\lambda_{\text{LED}}=\SI{310 \pm 10}{\nano\meter}$) (vi) are used to excite the sample. The deuterium lamp is mounted to an Acton VM-502 VUV-IR monochromator (vii), which is attached to the sample chamber, enabling excitation light at specific wavelengths (\SI{20}{\nano\meter} width). The deuterium lamp degrades in light output over time\cite[][p. 48]{AndiMA}. Therefore, a VUV-sensitive McPherson photodiode (viii) is used to monitor the output of the deuterium lamp.
The LED is operated by a constant current LED driver and is mounted on the cryo-shield.\par
For light collection, a \SI{5}{\milli\meter} diameter collimating lens directs some of the shifted light to an optical fiber, guiding it to an OceanOptics QE65000 UV-IR spectrometer (ix)\cite{AndiMA}. The lens is fixed perpendicularly to the excitation beam.\footnote{Conservatively estimated from the geometry of the setup the lens collects less than \SI{1}{\percent} of the emitted light.} By rotating the sample holder, the angle of incidence and detection can be adjusted between \SI{15}{\degree} and \SI{75}{\degree} in \SI{15}{\degree} steps. With the setup, measuring the relative wavelength shifting efficiency and emission spectra at different temperatures or comparing the properties between different samples in the same configuration is possible.\par
To prevent attenuation of VUV light on air, the entire system is evacuated to a vacuum of approximately \SI{e-7}{\milli\bar}. When cooling the sample, residual humidity can form a thin ice layer on the sample's surface. This surface effect has been shown to significantly reduce the amount of VUV light reaching the sample, affecting the measurement\cite{Neumeier}. To counter this effect, a two-step mitigation strategy is used. First, the cryo-shield and beaker filled with silica gel are cooled before the sample to adsorb the remaining humidity. Second, the UV LED is used as a normalization for the measurements as it is unaffected by the ice layer. To further limit the influence of the surface effect, the measurements with \SI{128}{\nano\meter} excitation were performed immediately after reaching the intended temperature\footnote{An additional temperature sensor attached to the sample itself during a calibration agreed in temperature to the other sensors.}.\par
\section{Characterization of wavelength shifters}
The WLSR sample used in this study is from the same material installed in LEGEND-200 WLSR. The sample comprises of a \SI{600}{\nano\meter} TPB layer coated on \SI{254}{\micro\meter} Tetratex\textregistered, backed by a \SI{50}{\micro\meter} copper foil. Further details on how the WLSR was coated and details of the commissioning can be found in\cite{SchwarzANNIMA}.\par
The PEN sample was cut from the same material installed in LEGEND-200. The material was formed using injection molding of commercially available PEN pellets from Teĳin-DuPont (TN-8065 SC). Additional information on the material's production and characterization is presented in \cite{PENproduction}. The amorphous PEN sample guides its emission light to its edges, resulting in a low signal-to-noise ratio in the VUV spectrofluorometer. To enhance the amount of detected light, the surface of the PEN sample was sanded using a 600-grit diamond sanding block, and a copper foil was placed for thermal coupling to the sample holder behind the sample.\footnote{The reproducibility of the surface quality after the manual sanding was investigated; the reflectivity of three produced samples varied by \SI{2}{\percent} between \SIrange{370}{600}{\nano\meter}.}\par
For each measurement, the sample was oriented at \SI{45}{\degree} to the excitation beam and the collimation lens. The obtained spectrum is corrected for the known response of the light detection system and the dark noise of the spectrometer. Furthermore, the data is refined by excluding noisy pixels. The stability of the deuterium lamp is monitored using the VUV-sensitive photodiode. Three sources of uncertainties are considered in the data analysis: 1) system instability, 2) uncertainty in the dark noise level, and 3) uncertainty in the response of the light detection system. To obtain the relative WLSE, the WLSR and PEN spectra are integrated between \SIrange{370}{550}{\nano\meter} and  \SIrange{370}{650}{\nano\meter}, respectively.\par
\section{Results and discussion}
\begin{figure}[htbp]
  \begin{subfigure}[t]{0.495\textwidth}
    \includegraphics[width=\textwidth]{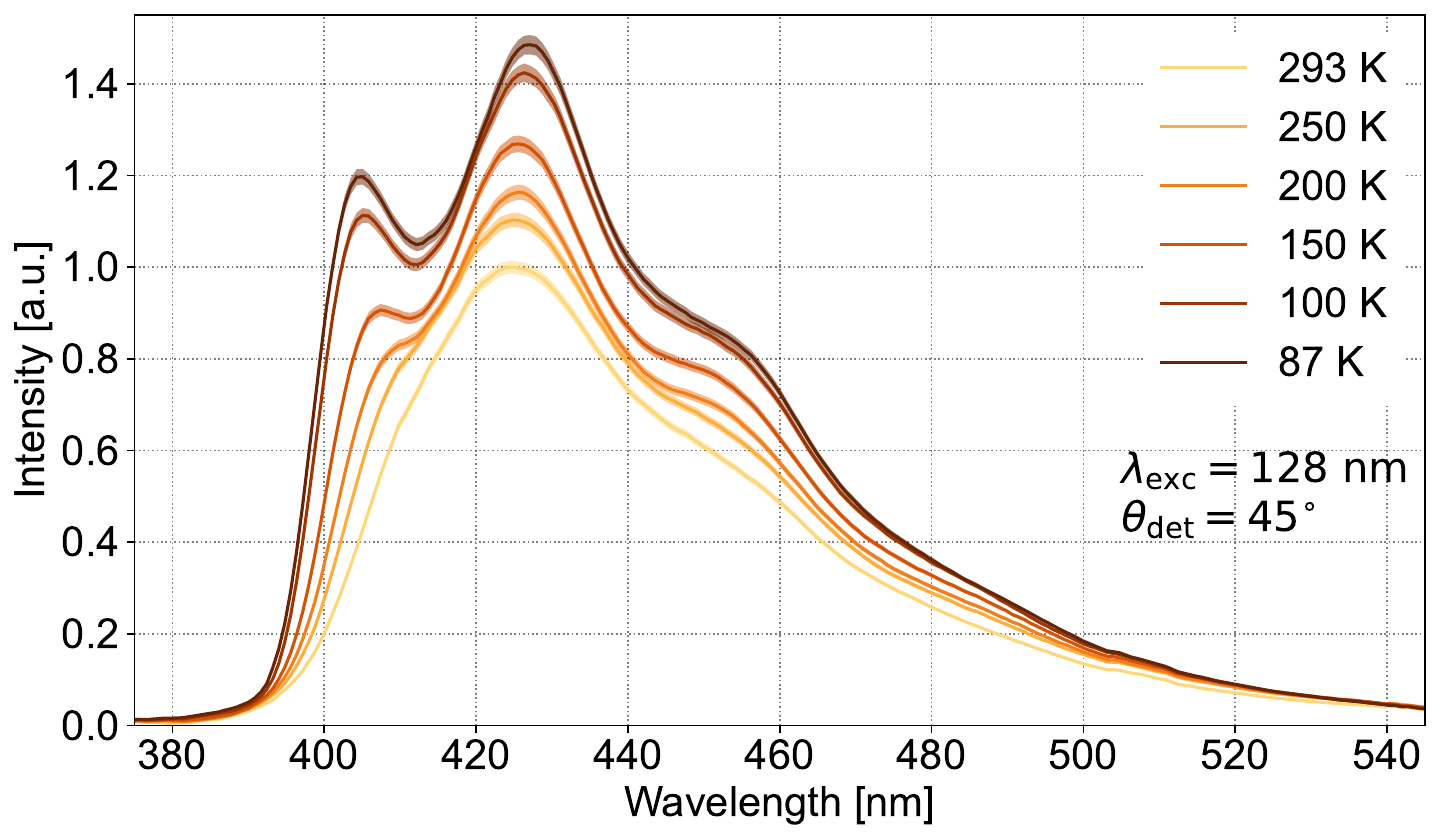}
    \centering
    \caption{at \SI{128}{\nano\meter} excitation.}
    \label{fig:WLSRSpectraLamp}
  \end{subfigure}
  \hfill
  \begin{subfigure}[t]{0.495\textwidth}
    \includegraphics[width=\textwidth]{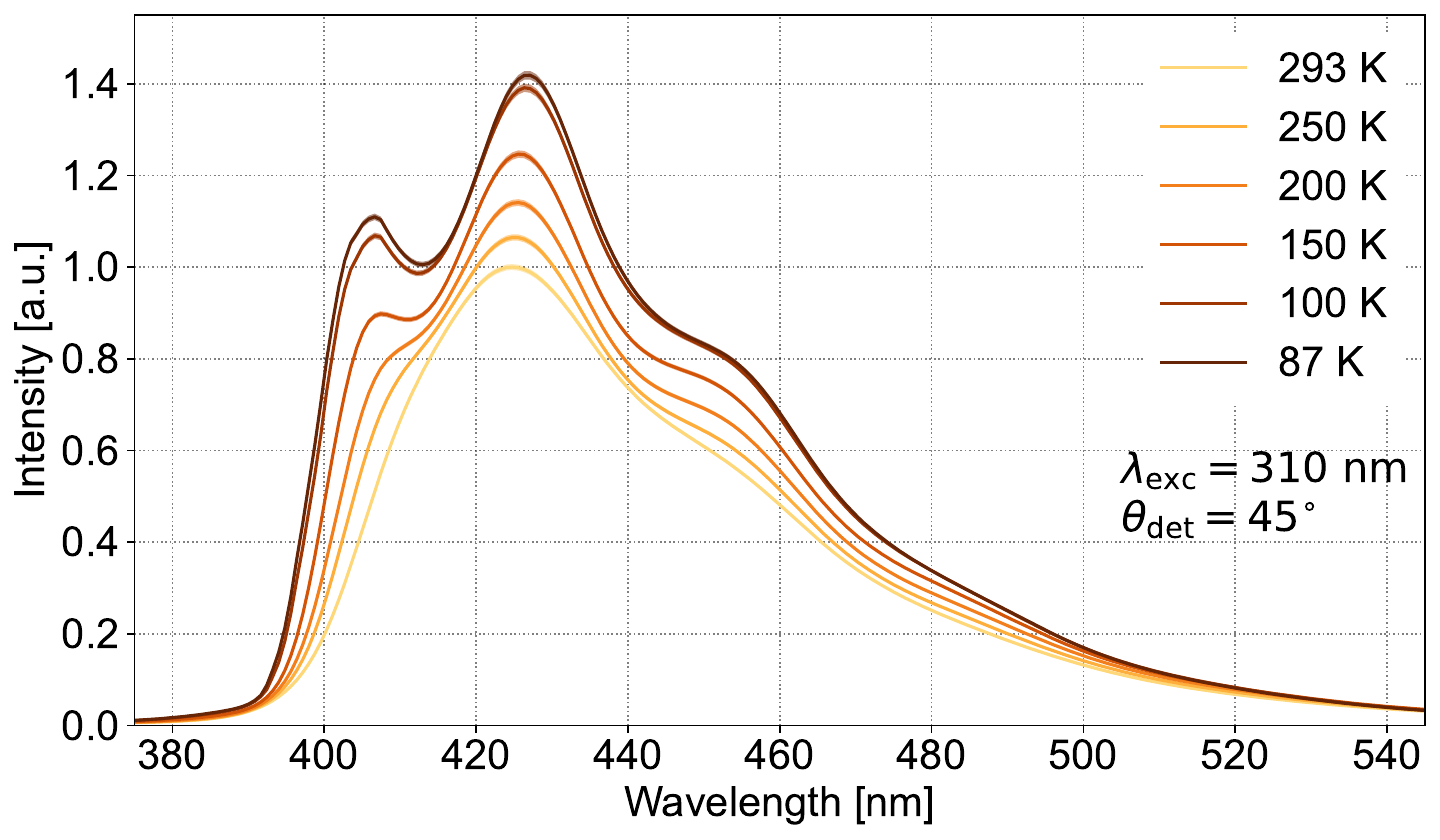}
    \centering
    \caption{at \SI{310}{\nano\meter} excitation.}
    \label{fig:WLSRSpectraLED}
  \end{subfigure}
  \caption{Emission spectra of the WLSR sample for temperatures between \SIrange{293}{87}{\kelvin}.}
  \label{fig:WLSRSpectra}
\end{figure}
The obtained spectra for the characterization of the WLSR sample are shown in \autoref{fig:WLSRSpectra}. The vibronic structures of the TPB molecule at \SIlist{405;450}{\nano\meter} emerge at lower temperatures. In \autoref{tab:WLSRPENLY}, the relative wavelength shifting efficiency (WLSE) of the cryogenic measurements compared to the measurement at room temperature (RT) is shown. We measured a strong increase in WLSE of \SI{54 \pm 5}{\percent} at LAr temperature (\SI{87}{\kelvin}) for excitation with LAr scintillation light wavelength (\SI{128}{\nano\meter}). The intrinsic quantum efficiency (QE) of TPB at RT for VUV excitation was measured by Benson \textit{et al.} to be \num{0.60 \pm 0.04}\cite{Benson}. Multiplied with the relative increase measured in this work, one gets a QE of approximately \num{0.92 \pm 0.08}, which is compatible with the QE of the LEGEND-200 WLSR obtained by Araujo \textit{et al.} of \num{0.85 \pm 0.05}(stat)\num{\pm 0.06}(syst.)\cite{Araujo}.\par
\begin{figure}[htbp]
  \begin{subfigure}[t]{0.495\textwidth}
    \includegraphics[width=\textwidth]{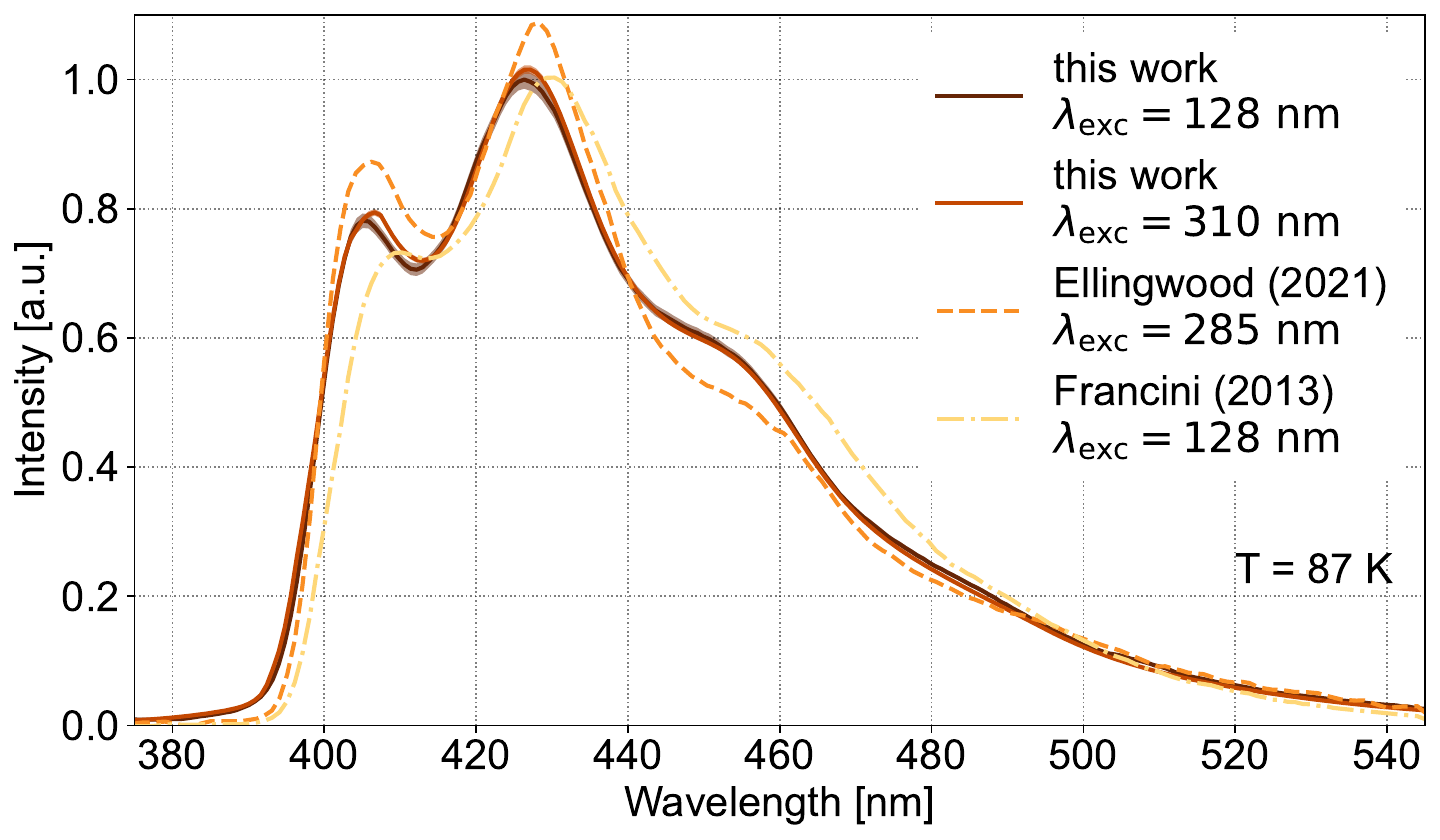}
    \centering
    \caption{Emission Spectra at \SI{87}{\kelvin}.}
    \label{fig:WLSRSpectraComparison}
  \end{subfigure}
  \hfill
  \begin{subfigure}[t]{0.495\textwidth}
    \includegraphics[width=\textwidth]{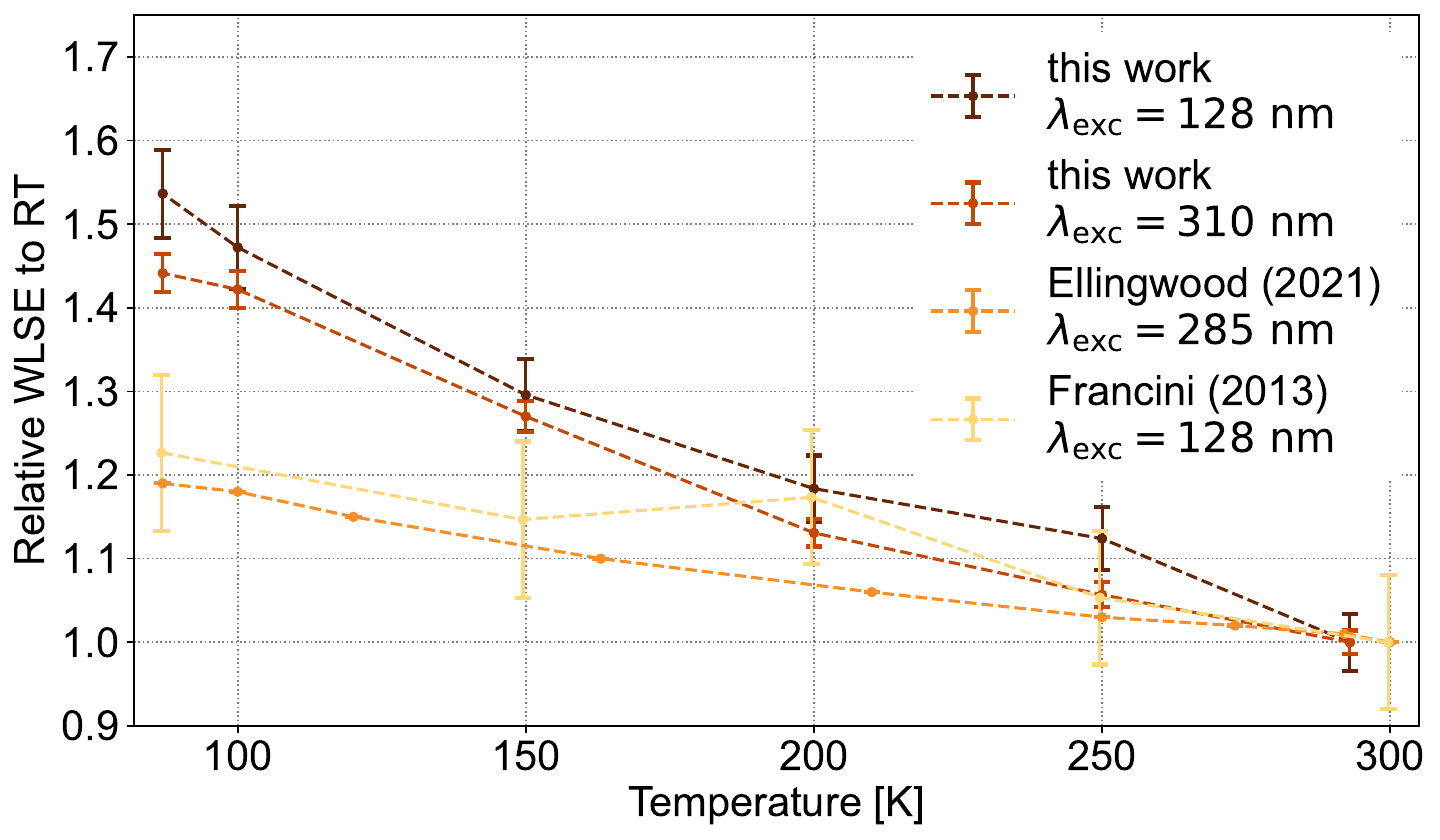}
    \centering
    \caption{Relative WLSE compared to RT for temperatures of \SIrange{300}{87}{\kelvin}.}
    \label{fig:WLSRLYComparison}
  \end{subfigure}
  \caption{Comparison to the results obtained by Ellingwood \textit{et al.}\cite{Ellingwood}, and Francini \textit{et al.}\cite{Francini}.}
  \end{figure}

\begin{table}[htbp]
\centering
\caption{Relative WLSE of the WLSR and PEN sample for cryogenic temperatures.}
\label{tab:WLSRPENLY}
\smallskip
\begin{tabular}{lcccc}
\hline
\multirow{3}{*}{Temperature [K]}&\multicolumn{4}{c}{Relative WLSE compared to RT [\%]}\\
&\multicolumn{2}{c}{WLSR}&\multicolumn{2}{c}{PEN}\\
&$\lambda_{\text{exc}}=\SI{128}{\nano\meter}$&$\lambda_{\text{exc}}=\SI{310}{\nano\meter}$&$\lambda_{\text{exc}}=\SI{128}{\nano\meter}$&$\lambda_{\text{exc}}=\SI{310}{\nano\meter}$\\
\hline
250 & \num{112 \pm 4} & \num{106 \pm 2} & \num{112 \pm 2} & \num{110 \pm 2}\\ 
200 & \num{118 \pm 4} & \num{113 \pm 2} & \num{123 \pm 3} & \num{122 \pm 2}\\
150 & \num{130 \pm 4} & \num{127 \pm 2} & \num{134 \pm 4} & \num{137 \pm 3}\\
100 & \num{147 \pm 5} & \num{142 \pm 2} & \num{134 \pm 4} & \num{152 \pm 2}\\
87  & \num{154 \pm 5} & \num{144 \pm 2} & \num{137 \pm 4} & \num{152 \pm 3}\\
\hline
\end{tabular}
\end{table}
Our results are compared with previous studies of TPB. Ellingwood \textit{et al.} investigated the emission of TPB excited by UV LED with a wavelength of $\lambda_{\text{exc}}=\SI{285}{\nano\meter}$. For their experiment, a \SI{1}{\micro\meter} thick TPB layer was coated on a \SI{5}{\milli\meter} thick PMMA substrate\cite{Ellingwood}. Francini \textit{et al.} characterized a TPB coating of approximately \SI{1.5}{\micro\meter} thickness on polymeric reflector (Vikuiti\texttrademark) deposited by vacuum evaporation\cite{Francini}. The emission spectrum at \SI{87}{\kelvin} is compared to these studies in \autoref{fig:WLSRSpectraComparison}.\par
The spectrum we detected is in good agreement with the results from \cite{Ellingwood} and \cite{Francini}. The minor differences in the spectra might originate in differences in the TPB vacuum evaporation procedure\cite[][p. 82]{AndiMA}, in differences in the reflectance/transmittance of the backing material, or the lack of calibrated response of the detection system used in \cite{Ellingwood} and \cite{Francini}.\par
In \autoref{fig:WLSRLYComparison}, the relative WLSE between RT and \SI{87}{\kelvin} is compared. The relative increase we measured both with \SI{128}{\nano\meter} and \SI{310}{\nano\meter} excitation exceeds the increase reported by \cite{Ellingwood} and \cite{Francini} by roughly a factor of two. The measurements by Francini \textit{et al.} might suffer strongly from the previously mentioned ice layer formation, lowering the increase in WLSE\cite{Neumeier}. Alternatively, the differences mentioned above in the TPB layer and the type of substrate may influence the performance at cold temperatures differently.\par
\begin{figure}[htbp]
  \begin{subfigure}[t]{0.32\textwidth}
    \includegraphics[width=\textwidth]{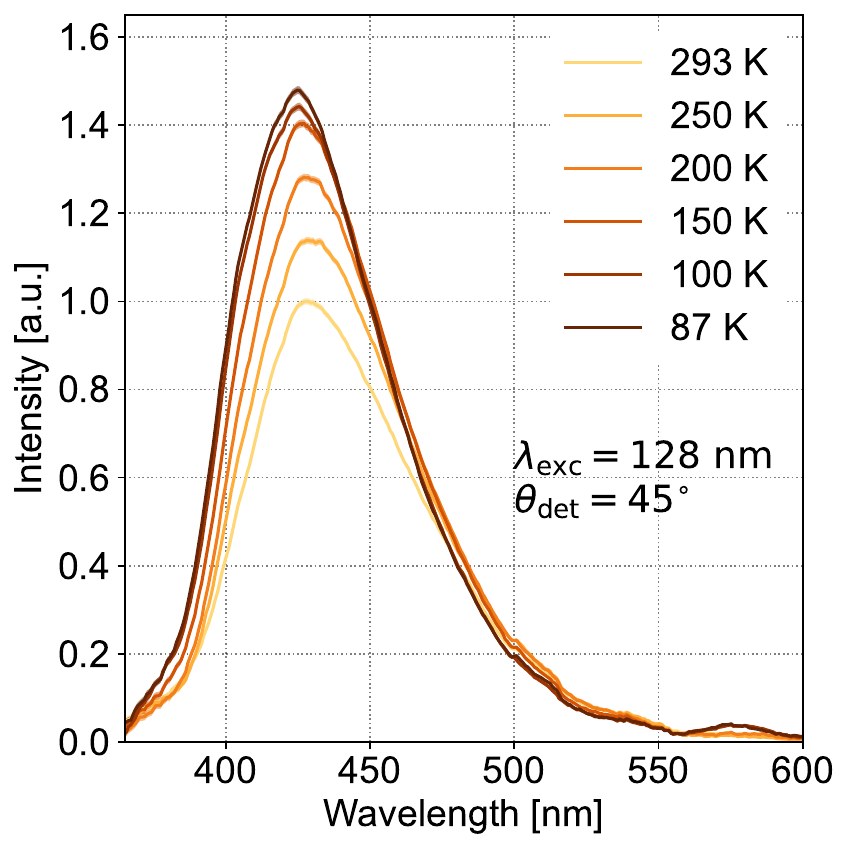}
    \centering
    \caption{at \SI{128}{\nano\meter} excitation.}
    \label{fig:PENSpectraLamp}
  \end{subfigure}
  \hfill
  \begin{subfigure}[t]{0.32\textwidth}
    \includegraphics[width=\textwidth]{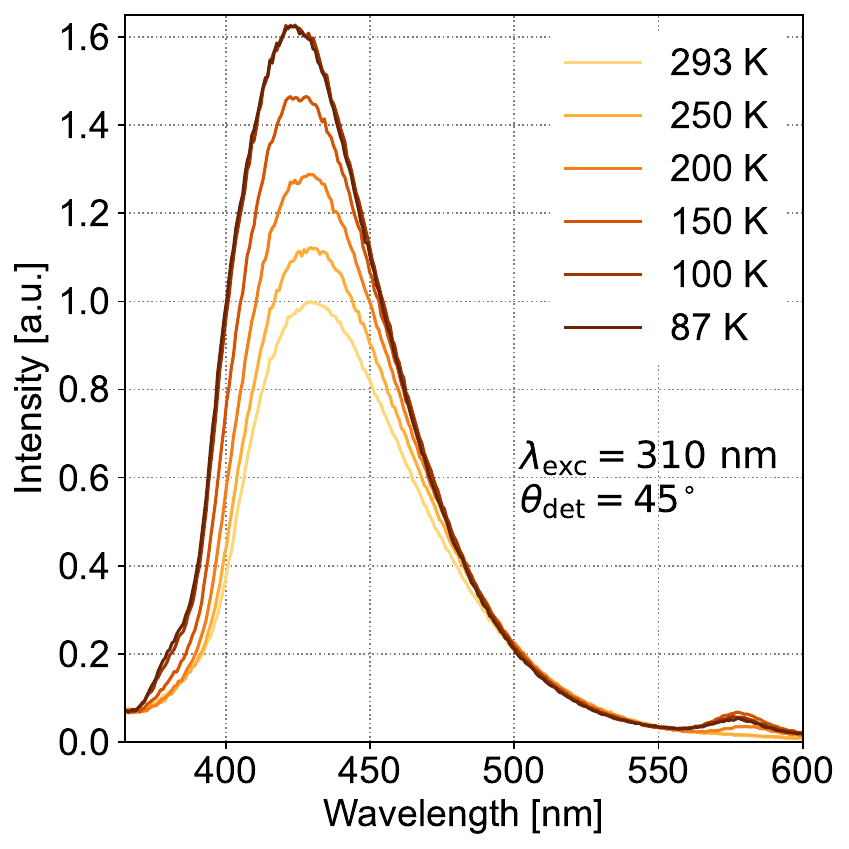}
    \centering
    \caption{at \SI{310}{\nano\meter} excitation.}
    \label{fig:PENSpectraLED}
  \end{subfigure}
  \hfill
  \begin{subfigure}[t]{0.32\textwidth}
  \includegraphics[width=\textwidth]{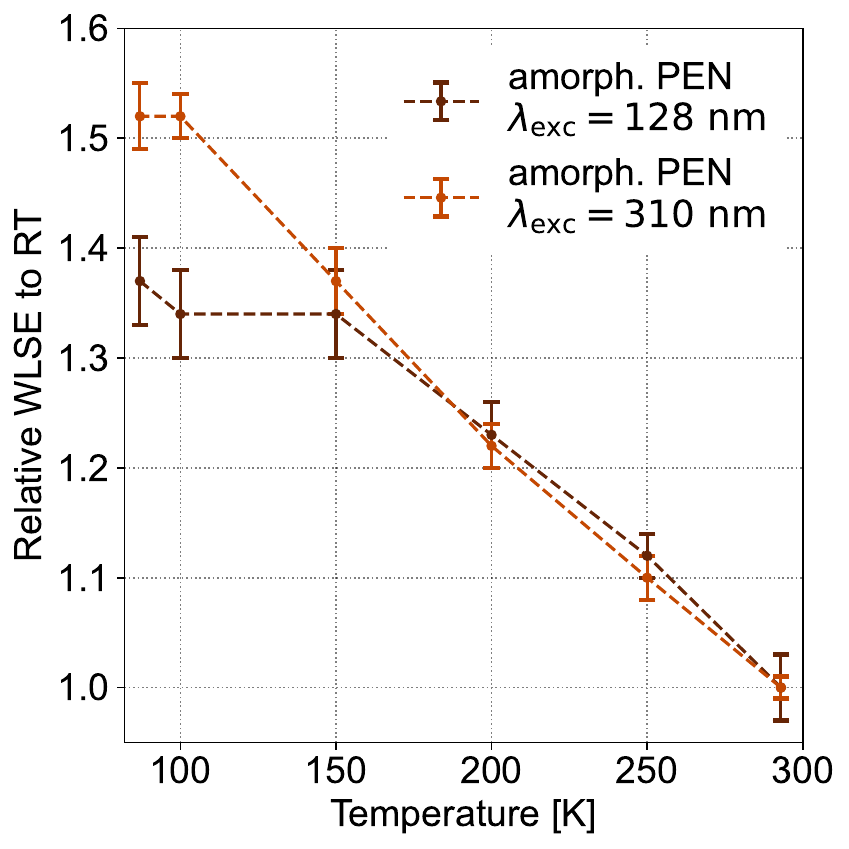}
    \centering
    \caption{relative WLSE}
    \label{fig:PENLY}
  \end{subfigure}
  \caption{Emission spectra and relative WLSE of the amorphous PEN sample for temperatures between \SIrange{293}{87}{\kelvin}.}
  \label{fig:PENResults}
\end{figure}
The emission spectra and change in WLSE of the amorphous PEN sample for temperatures between \SI{300}{\kelvin} and \SI{87}{\kelvin} are displayed in \autoref{fig:PENSpectraLamp} and \autoref{fig:PENSpectraLED}. A slight shift in the peak wavelength from \SI{430}{\nano\meter} for \SI{293}{\kelvin} to \SI{425}{\nano\meter} for \SI{87}{\kelvin} can be seen. Additionally, the monomeric emission at \SI{575}{\nano\meter} is minimally enhanced at cold temperatures. Both effects are in agreement with previous studies of PEN using \SI{300}{\nano\meter} excitation\cite{Mary}.\par
The relative increase in WLSE compared to \SI{293}{\kelvin} is presented in \autoref{fig:PENLY} and \autoref{tab:WLSRPENLY}. One can see a good agreement between the increase for both excitation wavelengths down to \SI{150}{\kelvin}. Below, the WLSE increases significantly less for the VUV excitation. The authors' hypothesis is that the surface effect, which increases in severity with cooling time, impacts the measurement with \SI{128}{\nano\meter} much stronger than in the WLSR measurement. Because of its larger mass compared to the WLSR sample and its poor thermal conductivity, the PEN sample needs more time to cool down to \SI{87}{\kelvin}. This is supported by the onset of the stagnation of the WLSE at \SI{150}{\kelvin}, which is predicted for \SI{e-7}{\milli\bar}\cite[][p. 25]{WaterIce}. In addition, measurements repeated at the final temperature of \SI{87}{\kelvin} with some delay show an even stronger degenerated WLSE. Hence, one can assume that the WLSE of PEN at LAr temperature (\SI{87}{\kelvin}) relative to \SI{293}{\kelvin} for excitation with LAr scintillation light wavelength (\SI{128}{\nano\meter}) is higher than the measured value of \SI{137 \pm 4}{\percent}.\par
\section{Conclusion}
We present the necessity and challenge of the characterization of WLS for the application in LAr-based experiments like LEGEND-200. For the characterization, we commissioned a VUV spectrofluorometer capable of measuring the emission spectrum and relative WLSE of wavelength shifting samples for excitation with LAr scintillation wavelength (\SI{128}{\nano\meter}) at temperatures from RT down to LAr temperature (\SI{87}{\kelvin}). For the TPB-based wavelength shifting reflector featured in LEGEND-200, the WLSE at \SI{87}{\kelvin} was found to be increased by \SI{54 \pm 5}{\percent} compared to RT. The WLSE of PEN, used in LEGEND-200 for the optically active detector holders, increased by at least \SI{37 \pm 4}{\percent} by cooling from \SIrange{293}{87}{\kelvin}. For excitation with \SI{310}{\nano\meter} light a respective increase by \SI{52 \pm 4}{\percent} was observed. To the knowledge of the authors, this work is the first characterization of the emission spectrum and relative WLSE of amorphous PEN at cryogenic temperatures with VUV and UV excitation.\par
\acknowledgments
This research is supported by the DFG through the Excellence Cluster ORIGINS EXC 2094 – 390783311 and the SFB1258.
\bibliographystyle{JHEP}
\bibliography{biblio.bib}
\end{document}